\documentclass[aps,prl,twocolumn,showpacs,preprintnumbers,amsmath,amssymb,superscriptaddress,floatfix]{revtex4}

\usepackage{graphicx}
\usepackage{dcolumn}
\usepackage{bm}
\newcommand{\nc}{\newcommand}
\nc{\be}{\begin{equation}}
\nc{\ee}{\end{equation}}
\nc{\bea}{\begin{eqnarray}}
\nc{\eea}{\end{eqnarray}}
\nc{\bean}{\begin{eqnarray*}}
\nc{\eean}{\end{eqnarray*}}
\nc{\mb}{\mbox}
\nc{\rnc}{\renewcommand}
\nc{\vk}{\mb{\bf k}}
\nc{\vp}{\mb{\bf p}}
\nc{\vn}{\mb{\bf n}}
\nc{\vq}{\mb{\bf q}}
\nc{\rr}{\mb{\bf r}}
\nc{\vz}{\hat {\mb{\bf z}}}
\nc{\vj}{\mb{\boldmath$j$}}
\nc{\vg}{\mb{\boldmath$g$}}
\nc{\x}{\mb{\boldmath$x$}}
\nc{\A}{\mb{\boldmath$A$}}
\nc{\va}{\mb{\boldmath$a$}}
\nc{\vs}{\mb{\boldmath$\sigma$}}
\nc{\vpi}{\mb{\boldmath$\pi$}}
\nc{\nab}{\nabla}
\nc{\X}{\sf x}

\begin{document}

\title{Chirality and Correlations in Graphene}

\author{Yafis Barlas}
\affiliation{Department of Physics, The University of Texas at Austin, Austin Texas 78712}
\author{T. Pereg-Barnea}
\affiliation{Department of Physics, The University of Texas at Austin, Austin Texas 78712}
\author{Marco Polini}
\affiliation{NEST-CNR-INFM and Scuola Normale Superiore, I-56126 Pisa, Italy}
\author{Reza Asgari}
\affiliation{Institute for Studies in Theoretical Physics and Mathematics, Tehran 19395-5531, Iran}
\author{A.H. MacDonald}
\affiliation{Department of Physics, The University of Texas at Austin, Austin Texas 78712}

\begin{abstract}
Graphene is described at low-energy by a massless Dirac equation whose eigenstates
have definite chirality.  We show that the tendency of Coulomb
interactions in lightly doped graphene to favor states with larger net chirality leads to suppressed
spin and charge susceptibilities.  Our conclusions are based on an evaluation of graphene's exchange
and random-phase-approximation (RPA) correlation energies.  The suppression is a consequence of
the quasiparticle chirality switch which enhances quasiparticle velocities near the Dirac point.
\end{abstract}

\maketitle

\noindent
{\em Introduction}---
Graphene can be described at low-energies by a massless Dirac-fermion (MDF)
model~\cite{slonczewski} with chiral quasiparticles that are responsible for
a number of unusual properties~\cite{earlytheory}.  The recent experimental realization of
single-layer graphene sheets~\cite{geim}
has made it possible to confirm some theoretical expectations, notably
an unusual half-quantized quantum Hall effect~\cite{dmqhe,graphene_qhe1,graphene_qhe2}.
In this Letter we show that quasiparticle chirality in weakly doped
graphene layers also leads to a peculiar suppression of the charge and spin susceptibilities.
We predict that both quantities are suppressed by approximately $15 \%$ in
current samples and that the suppression will be larger if uniform
samples with much lower densities can be realized.
At a qualitative level, these effects arise from an interaction energy preference
for MDF states with larger chiral polarization.
Our conclusions are based on an evaluation
of the exchange and RPA correlation energies of
uniform spin-polarized MDF systems with Coulomb interactions.

\noindent
{\em RPA Theory of Graphene}--- We study the following MDF model Hamiltonian (with $\hbar=1$),
\begin{equation}\label{eq:hamiltonian}
{\hat {\cal H}} = v\sum_{{\bf k},\alpha=\uparrow, \downarrow} {\hat{\psi}}^{\alpha\dagger }_{{\bf k}}
[\tau^{3}\otimes ({\bm \sigma} \cdot {\bf k})]{\hat \psi}_{{\bf k}}^{\alpha}
+\frac{1}{2S}\sum_{\tiny \begin{array}{c}{\bf q}\neq 0 \\ \alpha,\alpha'=\uparrow,\downarrow\end{array}}v_q {\hat n}_{{\bf q}}^{\alpha} {\hat n}_{-{\bf q}}^{\alpha'}\,,
\end{equation}
where $\tau^{3}$ is a Pauli matrix that acts on the two-degenerate ($K,K'$) valleys, ${\bf k}$ is a two-dimensional (2D) vector measured from the $K$
and $K'$ points, $\sigma^{1}$ and $\sigma^{2} $ are Pauli matrices that act on graphene's
pseudospin $(A,B)$ degrees of freedom, 
$S$ is the sample area, ${\hat n}_{\bf q}^\alpha = \sum_{{\bf k}} {\hat{\psi}}^{\dagger\alpha}_{{\bf k}-{\bf q}} {\hat \psi}_{{\bf k}}^{\alpha}$ is the single-spin density operator, and
$v_q=2\pi e^2/(\epsilon q)$ is the 2D Fourier transform of the Coulomb interaction potential $e^2/(\epsilon r)$.
In Eq.~(\ref{eq:hamiltonian}) the field operator ${\hat \psi}^{\alpha}_{{\bf k}} = ({\hat \Psi}^{\alpha}_{K,A},{\hat \Psi}^{\alpha}_{K,B},{\hat \Psi}^{\alpha}_{K',B},{\hat \Psi}^{\alpha}_{K',A})$ is a four-component spinor. In this work chirality refers to the eigenvalues $(s,s'=\pm)$ of ${\bm \sigma} \cdot {\bf k}/|{\bf k}|$.

The model (\ref{eq:hamiltonian}) requires an ultraviolet cutoff, as we discuss below.
To evaluate the interaction energy we follow a familiar strategy~\cite{Giuliani_and_Vignale} by combining
a coupling constant integration expression for the interaction energy valid for uniform continuum models,
\begin{equation}\label{eq:energy}
E_{\rm int} = \frac{N}{2}\int_0^{1}d\lambda
\int \frac{d^2 {\bf q}}{(2\pi)^2}~v_q~[S^{(\lambda)}(q)-1]\,,
\end{equation}
with a fluctuation-dissipation-theorem (FDT) expression~\cite{Giuliani_and_Vignale} for the static structure factor,
\begin{equation}
\label{eq:structurefactor}
S^{(\lambda)}(q) = -\frac{1}{\pi n}\int_0^{+\infty}
d\Omega ~\chi^{(\lambda)}_{\rho\rho}({\bf q},i\Omega)\,,
\end{equation}
where $n$ is the total electron density. This form of the FDT theorem takes advantage of the smooth behavior of the density-density response
function along the imaginary axis $ \chi^{(\lambda)}_{\rho\rho}({\bf q},i\Omega)$.  The RPA approximation for
the interaction energy then follows from the RPA approximation for $\chi$:
\begin{equation}\label{eq:chi_RPA}
\chi^{(\lambda)}_{\rho\rho}({\bf q},i\Omega) =
\frac{\chi^{(0)}({\bf q},i\Omega)}{1-\lambda v_q\chi^{(0)}({\bf q},i\Omega)}
\end{equation}
where $\chi^{(0)}({\bf q},i\Omega)$ is the non-interacting density-density response-function.
We have derived the following compact expression for
the $\chi^{(0)}$ contribution for an individual MDF model channel~\cite{mdfchi0theory}:
\begin{widetext}
\begin{equation}
\label{eq:final_result}
\chi^{\rm MDF}({\bf q},i\Omega)=-\frac{q^2}{16\sqrt{\Omega^2+v^2 q^2}}
-\frac{\varepsilon^{\rm c}_{\rm F}}{2\pi v^2}
+\frac{q^2}{8\pi\sqrt{\Omega^2+v^2 q^2}}\Re e\left[\sin^{-1}{\left(\frac{2\varepsilon^{\rm c}_{\rm F}+i\Omega}{vq}\right)}+
\left(\frac{2\varepsilon^{\rm c}_{\rm F}+i\Omega}{vq}\right)\sqrt{1-\left(\frac{2\varepsilon^{\rm c}_{\rm F}+i\Omega}{vq}\right)^2}\right]\,.
\end{equation}
In Eq.~(\ref{eq:final_result}) $\varepsilon^{\rm c}_{\rm F} = v k^{\rm c}_{\rm F}$ where $k^{\rm c}_{\rm F}$ is the channel Fermi momentum.
$\chi^{(0)}$ in Eq.~(\ref{eq:chi_RPA})
is constructed by summing the channel
response function ($\chi^{\rm MDF}$) over valley and spin with appropriate $\varepsilon^{\rm c}_{\rm F}$ values.
For a spin- and valley-unpolarized
system $\chi^{(0)} = g \chi^{\rm MDF}$ [with $k^{\rm c}_{\rm F}\to k_{\rm F}=(4\pi n/g)^{1/2}$]
where $g = g_{\rm s} g_{\rm v}=4$ accounts for spin and valley degeneracy.
\end{widetext}

\begin{figure}[t]
\begin{center}
\includegraphics[width=1.0\linewidth]{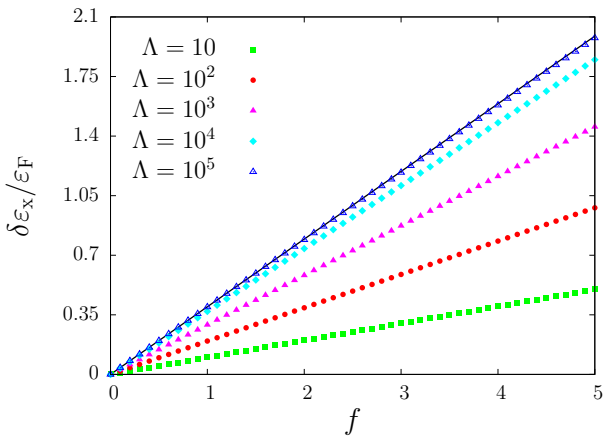}
\caption{(Color online) Cut-off $\Lambda$ and coupling constant $f$ dependence of the regularized exchange energy $\delta \varepsilon_{\rm x}$ in units of the Fermi energy $\varepsilon_{\rm F}$. The black solid line corresponds to the highest value of the cut-off that we have chosen, $\Lambda=2.7\times 10^5$.}
\label{fig:one}
\end{center}
\end{figure}

The energy constructed by combining Eqs.~(\ref{eq:energy})-(\ref{eq:final_result}) is clearly divergent since $\chi^{(0)}$
increases with $q$ at large $q$ and falls only like $\Omega^{-1}$ at large $\Omega$.
The divergence is expected since the energy calculated in this way
includes the interaction energy of the model's infinite sea of negative energy particles.
The MDF model can be expected to describe only changes in energy with density and spin-density
at small $\varepsilon^{\rm c}_{\rm F}$ values.  For definiteness we choose the total energy of undoped graphene
($\varepsilon^{\rm c}_{\rm F}=0$ in all channels) as our zero of energy.  For pedagogical  and numerical reasons it is
also helpful to separate the contribution that is first order in $e^2$, the exchange energy,
from the higher order contributions conventionally referred to in electron gas theory as the correlation energy.
Using Eqs.~(\ref{eq:energy})-(\ref{eq:final_result}) we find that for unpolarized doped graphene
the excess exchange energy per excess electron is
\begin{widetext}
\begin{equation}\label{eq:exchange_regularized}
\delta \varepsilon_{\rm x} = -\frac{1}{2\pi n}\int \frac{d^2 {\bf q}}{(2\pi)^2}~v_q \int_0^{+\infty}
d\Omega ~\left[\chi^{(0)}({\bf q},i\Omega)-\left.\chi^{(0)}({\bf q},i\Omega)\right|_{\varepsilon_{\rm F}=0}\right]
\equiv -\frac{1}{2\pi n}\int \frac{d^2 {\bf q}}{(2\pi)^2}~v_q \int_0^{+\infty}
d\Omega~\delta \chi^{(0)}({\bf q},i\Omega)\,,
\end{equation}
and that the corresponding correlation energy is
\begin{equation}\label{eq:regularization}
\delta \varepsilon^{\rm RPA}_{\rm c}= \frac{1}{2\pi n}\int \frac{d^2 {\bf q}}{(2\pi)^2}\int_0^{+\infty}d\Omega\left\{v_q \delta \chi^{(0)}({\bf q},i\Omega)+
\ln{\left[\frac{1-v_q\chi^{(0)}({\bf q},i\Omega)}{1-v_q\left.\chi^{(0)}({\bf q},i\Omega)\right|_{\varepsilon_{\rm F}=0}}\right]}\right\}\,.
\end{equation}
\end{widetext}
With this regularization the $\Omega$ integrals are finite and the $q$ integrals have logarithmic ultraviolet
divergences.  The remaining divergences are physical and follow from the interaction between
electrons near the Fermi energy and electrons very far from the Fermi
energy as we discuss at length later.  The best we can do in using the MDF model to
make predictions relevant to graphene sheets is to introduce an ultraviolet cutoff for the
wavevector integrals, $k_{\rm c}$.  $k_{\rm c}$ should be assigned a value corresponding to the
wavevector range over which the MDF model describes graphene.  Based on this criterion~\cite{minso}
we estimate that $k_{\rm c} \sim 1/a$ where $a \sim 0.246$ nm is graphene's lattice constant.
The MDF is useful when $k_{\rm c}$ is much larger than $k_{\rm F}$ in all channels.

\begin{figure}[t]
\begin{center}
\includegraphics[width=1.0\linewidth]{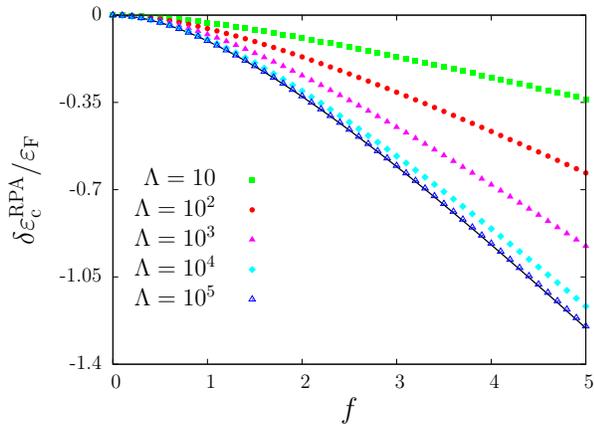}
\caption{(Color online) Cut-off and coupling constant dependence of the regularized correlation energy $\delta \varepsilon^{\rm RPA}_{\rm c}$ in units of the Fermi energy $\varepsilon_{\rm F}$. \label{fig:two}}
\end{center}
\end{figure}

With this regularization the properties of graphene's MDF model depend on the dimensionless coupling constant (restoring $\hbar$)
\begin{equation}
\label{eq:coupling}
f \equiv g~\frac{e^2}{\epsilon v \hbar}=\frac{g}{\epsilon}\frac{c}{v} \alpha\,,
\end{equation}
and on $\Lambda = k_{\rm c}/k_{\rm F}$.  In Eq.~(\ref{eq:coupling}), $c \sim 300 v$ is the speed
of light, $\alpha \approx 1/137$ is the fine structure constant, and $\epsilon$ depends on the dielectric
environment of the graphene layer.  In typical circumstances $f \sim 2$.  $\Lambda$ is $\sim 10$
in the most heavily doped samples studied experimentally and can in principle
be arbitrarily large in lightly doped systems.
We expect, however, that many of the electronic properties of graphene layers will
be dominated by disorder when the doping is extremely light.

In Fig.~\ref{fig:one} and Fig.~\ref{fig:two}
we plot the exchange and correlation energies of the graphene MDF model as a function of $f$ for
a range of $\Lambda$ values.  Note that both $\delta \varepsilon_{\rm x}$ and
$\delta \varepsilon^{\rm RPA}_{\rm c}$ have the same density dependence as
$\varepsilon_{\rm F} \propto n^{1/2}$ apart from the weak dependence on $\Lambda$.  The exchange energy
is positive because our regularization procedure implicitly selects the chemical potential
of undoped graphene as the zero of energy; doping either occupies quasiparticle states with
positive energies or empties quasiparticles with negative energies.  Note that including the
RPA correlation energy weakens the $\Lambda$ dependence so that the exchange energy per electron
scales more accurately with $\varepsilon_{\rm F}$.
It is possible to analytically extract the asymptotic behavior of the exchange and correlation energies
at large $\Lambda$ by Laurent expanding the integrands of Eqs.~(\ref{eq:exchange_regularized})-(\ref{eq:regularization})
in $q$ and retaining only the $1/q$ terms:
\begin{equation}\label{eq:cut-off-exchange}
\delta \varepsilon_{\rm x}=\frac{1}{6g}f \varepsilon_{\rm F}\ln{(\Lambda)} + {\rm regular~terms}\,,
\end{equation}
and
\begin{equation}
\label{eq:cut-off-correlation}
\delta \varepsilon^{\rm RPA}_{\rm c}=-\frac{1}{6g}f^2 \xi(f)\varepsilon_{\rm F}\ln{(\Lambda)}+ {\rm regular~terms}
\end{equation}
where
\begin{equation}\label{eq:xi}
\xi(f)=4\int_0^{+\infty}
\frac{dx}{(1+x^2)^{2}(8\sqrt{1+x^2}+f\pi)}\,.
\end{equation}
[Note that $\xi(f=0)=1/3$ so that the exchange and correlation energies are comparable in size for typical $f$ values.]

\noindent
{\em Charge and Spin Susceptibilities}---
In an electron gas, the physical observables most directly related to the energy are the
$\Omega\rightarrow 0, q\rightarrow 0$ charge and spin-susceptibilities,
normally discussed in terms of dimensionless ratios between
non-interacting and interacting system values.
The charge susceptibility is the inverse of the thermodynamic
compressibility $\kappa$ of the system up to a factor of $n^2$.  For the MDF model of doped graphene
\begin{equation}
\label{eq:kappainv}
\frac{\kappa_0}{\kappa} = \frac{2n}{\varepsilon_{\rm F}} \frac{\partial^2 (n \delta\varepsilon_{\rm tot})}{\partial n^2}
= \frac{2n}{\varepsilon_{\rm F}} \frac{\partial \mu}{\partial n}\,,
\end{equation}
and
\begin{equation}
\label{eq:chiinv}
\frac{\chi_0}{\chi_{\rm S}}=
\frac{2}{\varepsilon_{\rm F}}\left.\frac{\partial^2 [\delta \varepsilon_{\rm tot}(\zeta)]}{\partial \zeta^2}\right|_{\zeta=0}\,,
\end{equation}
where $\delta\varepsilon_{\rm tot}$ includes band, exchange, and correlation contributions.
In Eq.~(\ref{eq:chiinv}) $\zeta \equiv (n_{\uparrow}-n_{\downarrow})/(n_{\uparrow}+n_{\downarrow})$, $\chi^{-1}_{\rm S}$ measures the stiffness of the system against changes in the density of electrons with spin $\uparrow$ and spin $\downarrow$. In a 2D electron systems the compressibility can be measured~\cite{eisensteincompress} capacitively.  We note that this type of measurement is less difficult when $\partial \mu/\partial n$ is large as it is in weakly-doped graphene.
In bulk electronic systems, the spin-susceptibility can usually be extracted successfully from
total magnetic susceptibility measurements, but these are likely to be challenging in the case of single-layer graphene.
In 2D electron systems, however, information about the spin-susceptibility can often~\cite{wfsdh} be extracted from weak-field magnetotransport experiments using a tilted magnetic field to distinguish
spin and orbital response.

\begin{figure}[t]
\begin{center}
\includegraphics[width=1.0\linewidth]{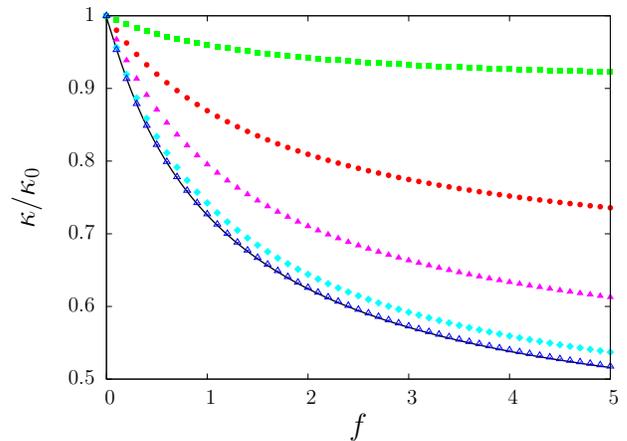}
\caption{(Color online) Cut-off and coupling constant dependence of $\kappa/\kappa_0$. The color coding is as in Figs.~\ref{fig:one}-\ref{fig:two}.}
\label{fig:three}
\end{center}
\end{figure}

\begin{figure}[t]
\begin{center}
\includegraphics[width=1.0\linewidth]{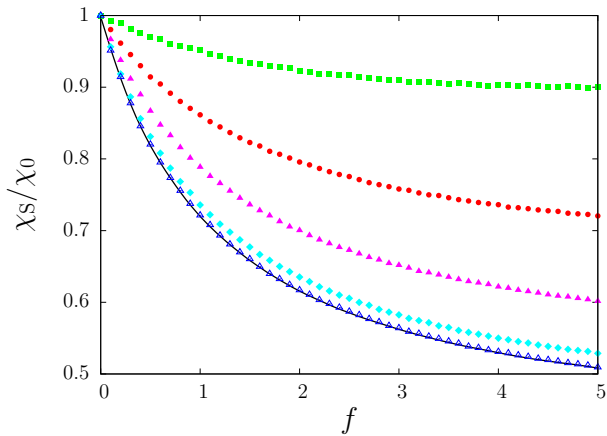}
\caption{(Color online) Cut-off and coupling constant dependence of the spin susceptibility $\chi_{\rm S}$. The color coding is as in Figs.~\ref{fig:one}-\ref{fig:two}.}
\label{fig:four}
\end{center}
\end{figure}

Our results for the charge and spin-susceptibilities are summarized in Fig.~\ref{fig:three} and Fig.~\ref{fig:four}.
For experiments performed over the density range over which properties
appear to be intrinsic in current samples ($\Lambda$ between $\sim 10$ and $\sim 40$),
these results predict compressibility and susceptibility suppression
(apparent quasiparticle velocity enhancement) by approximately $15\%$.
Both the sign of the interaction effect and the similarity of $\kappa$ and $\chi$
are in remarkable contrast with familiar electron gas behavior.
In 3D and 2D non-relativistic electron gases both are~\cite{Giuliani_and_Vignale,Guinea}
strongly enhanced by interactions, with the charge response diverging
at intermediate coupling and the spin response diverging at very strong coupling.

\noindent
{\em Discussion}---
The qualitative physics of ferromagnetism in metals is most transparent at the
Hartree-Fock (HF) level.  Similarly, the mechanism responsible for the unusual interaction
physics of weakly doped graphene becomes clear when the exchange energy is expressed in terms of HF theory quasiparticle
self-energies.  Correlations do however play an essential quantitative role.  For
doped graphene the contribution of an individual channel to the HF theory interaction energy is
\begin{eqnarray}
\delta\varepsilon_{\rm x}&=&-\frac{1}{2nS^2}
\sum_{s, s'} \sum_{{\bf k},{\bf k}'}
V_{s,s'}({\bf k},{\bf k}')\,\delta n_{{\bf k}s}\delta n_{{\bf k}'s'}\nonumber\\
&+&\frac{1}{n S}
\sum_{{\bf k},s}\Sigma^{(0)}_{{\bf k},s}\delta n_{{\bf k}s}
\label{eq:hfexchangeenergy}
\end{eqnarray}
where $s,s'=\pm$ are the chirality indices of the MDF bands ({\it i.e.} the eigenvalues of the chirality operator defined above),
\begin{equation}
\Sigma^{(0)}_{{\bf k},s} = -\frac{1}{S}\sum_{{\bf k}', s'}V_{ss'}({\bf k},{\bf k}')n^{(0)}_{s'}({\bf k}')
\end{equation}
is the HF self-energy of the {\it undoped} MDF model,
\begin{equation}
\label{eq:xme}
V_{s,s'}({\bf k},{\bf k}') = \frac{2 \pi e^2}{|{\bf k}-{\bf k}'|}  \; \left[ \frac{1+ s s' \cos(\theta_{{\bf k},{\bf k}'})}{2} \right]
\end{equation}
is the exchange matrix elements between band states $s',\bf{k'}$ and $s,\bf{k}$,
and $\theta_{{\bf k},{\bf k}'}$ is the angle between ${\bf k}$ and ${\bf k}'$.
For Coulomb interactions, the factor in square brackets on the right hand side of Eq.~(\ref{eq:xme})
tends to be larger between states in the same band, {\em i.e.} states with the same chirality.

The first term on the right-hand side of Eq.~(\ref{eq:hfexchangeenergy}) is similar to the
exchange energy of an ordinary 2D electron system.  Because it is negative and
increases with density, its contribution to the exchange energy is lowered when spins are unequally
populated.  If this was the only exchange energy contribution, the spin-susceptibility
and inverse compressibility would
be enhanced by interactions as usual.  The unusual behavior comes from the second term.
For weakly doped graphene it is sufficient to expand $\Sigma^{(0)}_{{\bf k},s}$ to first order
in $k$; $\Sigma^{(0)}_{{\bf k}=0,s}$ is a physically irrelevant constant that is
included in the chemical potential chosen as the zero of energy by our renormalization procedure.
Expanding to first order in $k$ gives the leading interaction contribution to the
velocity renormalization~\cite{velocityrenorm} of undoped graphene.
In agreement with previous work we find that for large $\Lambda$ and small $f$
\begin{equation}
v \to v \left[ 1 + \frac{f}{4 g} \ln(\Lambda) \right]\,.
\label{eq:vel}
\end{equation}
The physical origin of the velocity increase is the loss in exchange energy
on crossing the Dirac point from states that have the same chirality as
the occupied negative energy sea to states that have the opposite chirality.
It is easy to verify that this velocity renormalization is responsible for the
leading $\ln(\Lambda)$ terms in the exchange energy and in the exchange contributions
to $\kappa^{-1}$ and $\chi^{-1}_{\rm S}$.  The conventional
exchange energy contributes negatively to $\kappa^{-1}$ and $\chi^{-1}_{\rm S}$
and competes with the Dirac point velocity renormalization.

When correlations are included, the leading $\ln(\Lambda)$
contributions to the interaction energy and to $\kappa^{-1}$ and $\chi^{-1}_{\rm S}$,
still follow from the (now altered) undoped system quasiparticle velocity renormalization.
The enhanced quasiparticle velocity is tied to the Dirac point, {\em i.e.} to the
switch in chirality, and results in an interaction energy that tends to be lower when
the chemical potential is close to the Dirac point in all channels.
Fig.~\ref{fig:three} and Fig.~\ref{fig:four} illustrate RPA theory predictions
for experimentally observable consequences of the competition between
this interband effect and conventional intraband correlations.

\noindent
{\em Acknowledgment ---}
This work has been supported by the Welch Foundation, by the Natural Sciences and Engineering Research Council of Canada,
by the Department of Energy under grant DE-FG03-02ER45958, and by the National Science Foundation under grant DMR-0606489.
M.P. acknowledges useful discussions with F. Poloni.

\end{document}